\documentclass{iaus}
\usepackage{graphicx}

\newcommand{\lis}{$^7$Li/$^6$Li }
\newcommand{\bb}{$^{11}$B/$^{10}$B }

\title[Evolution of spallogenic nuclides] %% give here short title %%
{Origin of cosmic rays and evolution of spallogenic nuclides Li, Be and B}

\author[N. Prantzos]   %% give here short author list %%
{Nikos Prantzos}%%  \thanks{Present address: Fluid Mech Inc., 24 The Street, Lagos, Nigeria.},

\affiliation{Institut d' Astrophysique de Paris \\ 
98bis, Bd. Arago, 75014 Paris \\ 
email: {\tt prantzos@iap.fr} }

\pubyear{2010}
\volume{268}  %% insert here IAU Symposium No.
\jname{Light elements in the Universe}
\editors{C. Charbonnel, M. Tosi, F. Primas \& C. Chiappini, eds.}

\begin{document}

\maketitle

\begin{abstract}
A short overview is presented of current issues concerning the production
and evolution of Li, Be and B in the Milky Way. In particular, the observed ``primary-like" 
evolution  of Be is re-assessed in the light of a novel idea: it is argued that  
 Galactic Cosmic Rays  are accelerated from the wind material of {\it rotating} massive stars,
 hit by the forward shock of the subsequent supernova explosions. The pre-galactic 
 levels of both Li isotopes  remain controversial at present, making it difficult
to predict their Galactic evolution. A quantitative  estimate is provided of the contributions of
various candidate sources to the solar abundance of Li.
\end{abstract}

\firstsection % if your document starts with a section,

\section{Introduction}
The idea that the light and fragile elements Li, Be and B are
produced by the interaction  of the energetic nuclei
of galactic cosmic rays (CGR) with the nuclei of the interstellar medium
(ISM) was introduced 40 years ago (Reeves et al. 1970, Meneguzzi et al. 1971, hereafter MAR).
In those early works it was shown that, taking into account
the relevant cross-sections and with plausible assumptions about the
GCR properties - source composition, intensity and spectrum -
one may reproduce reasonably well the abundances of those light
elements observed in GCR and in  meteorites (pre-solar).

Among the required ingredients for such a calculation, the relevant spallation cross sections
of CNO nuclei are accurately  measured in the laboratory. The source composition and the equilibrium energy
spectrum of GCR are inferred from a combination of observations and models of GCR propagation
in the Milky Way (e.g. in the framework of the so-called ``leaky box" model).
Once the equilibrium spectra of GCR in the ISM are established, the calculation
of the resulting abundances of LiBeB is straightforward, at least to first order\footnote{The
full calculation should include production by spallation  of other primary and secondary 
nuclides, such as $^{13}$C; however, this has only second order effects.}. The production rate (s$^{-1}$)
of the abundance $Y_L=N_L/N_H$ (by number) of LiBeB nuclei is given by
\begin{equation}
\frac{dY_L}{dt} \ = \ F^{GCR}_{p,a}\sigma_{pa+CNO}Y^{ISM}_{CNO} \ +
                       F^{GCR}_{CNO}\sigma_{pa+CNO}Y^{ISM}_{p,a} P_L\ +
                       F^{GCR}_{a}\sigma_{a+a}Y^{ISM}_{a} P_L
\end{equation}
where: $F$ (cm$^{-2}$ s$^{-1}$) is the average GCR flux of protons, alphas or CNO, $Y$ the abundances 
by number of those nuclei
in the ISM, and $\sigma$ (cm$^2$) is the average (over the equilibrium energy spectrum of GCR) 
cross-section for the corresponding spallation reactions producing LiBeB.
The first term in the right hand member of this 
 equation (fast protons and alphas hitting CNO nuclei of the ISM) is
known as the ``direct" term, the second one (fast CNO nuclei being fragmented on ISM protons and 
alphas) is the ``reverse" term and the last one involves ``spallation-fusion" 
reactions, concerning only the Li isotopes. $P_L$ is the probability that nuclide $L$ (produced at high energy) 
will be thermalized and remain in the ISM (see, e.g. Prantzos 2006). 
Obviously, the GCR flux term $F^{GCR}_{CNO} \propto Y^{GCR}_{CNO}$
is proportional to the abundances of CNO nuclei in GCR, a fact of paramount importance for the evolution
of Be and B (see next sections).

Substituting appropriate values for GCR fluxes ($F^{GCR}_p\sim$10 p cm$^{-2}$ s$^{-1}$ 
 for protons
and scaled values for other GCR nuclei), for the corresponding cross sections (averaged
over the GCR equilibrium spectrum $\sigma_{p,a+CNO\longrightarrow Be}\sim$10$^{-26}$ cm$^{-2}$) and
for ISM abundances $Y_{CNO}\sim$10$^{-3}$, and integrating for $\Delta t \sim$10$^{10}$ yr, one
finds $Y_{Be}\sim$2 10$^{-11}$, i.e. approximately the meteoritic Be value. Satisfactory results
are also obtained for $^6$Li and $^{10}$B.

Two problems were identified with the GCR production, compared
to meteoritic composition: 
the \lis ratio ($\sim$2 in GCR, but $\sim$12 in meteorites) and the \bb
ratio ($\sim$2.5 in GCR, but $\sim$4 in meteorites). It was then suggested in MAR that supplementary
sources are needed for $^7$Li and $^{11}$B. Modern solutions to
those problems involve {\it stellar} production of $\sim$60\% of
$^7$Li (in the hot envelopes of AGB stars and/or novae, see Sec. 7) and of $\sim$40\% 
of $^{11}$B (through $\nu$-induced spallation of $^{12}$C in SN, see Sec. 5). In both
cases, however, uncertainties in the yields are such that observations
are used to  constrain the yields of the candidate sources
rather than to confirm the validity of the scenario.

\begin{figure}[t!]
\begin{center}
\includegraphics[width=0.99\textwidth]{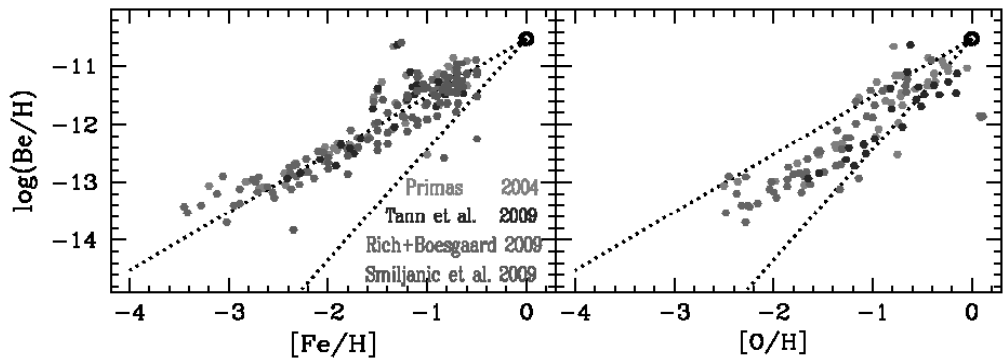}
\caption[]{ Observations of Be vs. Fe ({\it left}) and vs. O ({\it right}). In both panels, 
dotted lines indicate slopes of 1 (primary) and 2 (secondary). Be clearly behaves  
as a primary vs. Fe, whereas there is more scatter in the data vs. O.
}
\label{eps1}
\end{center}
\end{figure}

\begin{figure}[t!]
\begin{center}
\includegraphics[angle=-90,width=0.7\textwidth]{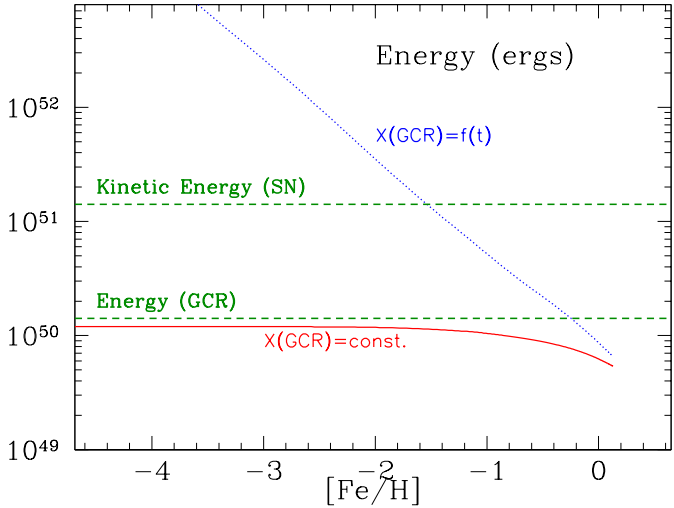}
\caption[]{Energy input required from energetic particles accelerated by one CCSN in order
to produce a given mass of Be, such as
to have [Be/Fe]=0 (solar), assuming that a core collapse SN produces, on average, 0.1 M$_{\odot}$
of Fe. {\it Solid} curve corresponds to the case of a constant composition for GCR, {\it dotted} curve corresponds
to a time variable composition, following the one of the ISM. In the former case, the required energy
is approximately equal to the energy imparted to energetic particles by supernovae, namely $\sim$0.1
of their kinetic energy of  $\sim$1.5 10$^{51}$ ergs; in the latter case, the energy required to
keep [Be/Fe]=0 becomes much larger than the total kinetic energy of a CCSN
for metallicities [Fe/H]$\leq$-1.6.}
\label{eps1}
\end{center}
\end{figure}

\section{Primary Be: the problem}

Observations of halo stars in the 90s revealed a linear relationship
between Be/H and Fe/H (Gilmore et al. 1991, Ryan et al. 1992) 
as well as between B/H and Fe/H (Duncan et al. 1992). 
That was unexpected, since Be and B
were thought to be produced as {\it secondaries}, by spallation of the 
increasingly abundant CNO nuclei. Indeed, the first two terms in Eq. 1.1
were thought to evolve in the same way with time (or metallicity), since the
composition of GCR Y$_{CNO}^{GCR}$ was supposed to evolve in step with the one of the 
ISM Y$_{CNO}^{ISM}$. Only the Li isotopes, produced at low
metallicities mostly by $\alpha+\alpha$  reactions  were thought to be produced  as 
primaries (Steigman and Walker
1992) . The only way to produce primary Be  is
by assuming that GCR have always the same CNO content, as suggested in Duncan et al. (1992).
Other efforts to enhance the early production of Be, by e.g. invoking a better confinement - 
and thus, higher fluxes - of GCR in the early Galaxy (Prantzos et al. 1993) 
%or a higher O/Fe ratio at low metallicities (Fields and Olive 1999), 
failed. The reason for that failure
was clearly revealed by the ``energetics argument"
put forward by Ramaty et al. (1997): if SN are the main source of GCR energy,
there is a limit to the amount of light elements produced per SN, which depends
on GCR and ISM composition. If the metal content of {\it both} ISM and GCR is
low, there is simply not enough energy
in GCR to keep the Be yields constant (Fig. 2)\footnote{For reasons unknown to the author, the
energetics argument was obviously not understood by many prolific
researchers in the field in the late 90ies.}.
Since the ISM metallicity certainly increases with time, the ``direct" component in Eq. 1.1
produces only secondary LiBeB. The only possibility
to have $\sim$constant LiBeB yields is by assuming that the ``reverse" component is primary,
i.e. that GCR have a $\sim$constant metallicity. This has profound implications for our 
understanding of the GCR origin. It should be noted that before those Be and B observations,
no one would have the idea to ask ``what was the GCR composition in the early Galaxy?".

\begin{figure}[t!]
\begin{center}
\includegraphics[width=0.9\textwidth]{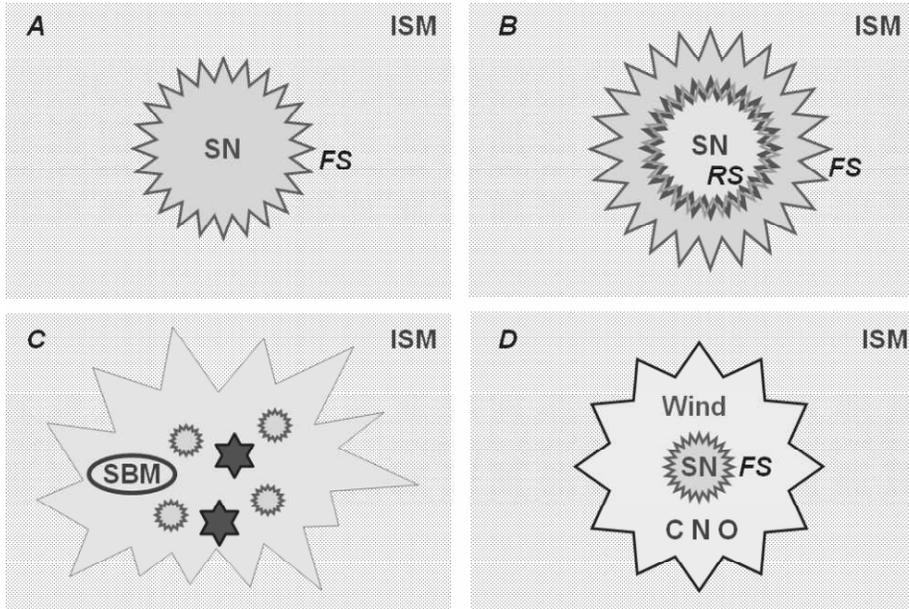}
\caption[]{ Scenarios for the origin of Galactic cosmic rays (GCR). 
{\bf {\it A}}:  GCR originate from the interstellar medium (ISM) and are accelerated from the
forward shock (FS) of supernovae (SN).
{\bf {\it B}}:  GCR originate from the interior of supernovae and are accelerated by the 
reverse shock (RS), propagating inwards.
{\bf {\it C}}:  GCR originate from superbubble material (SBM), enriched by the metals 
ejected by supernovae and massive star winds; they are accelerated by the forward shocks of 
supernovae {\it and} stellar winds.
{\bf {\it D}}:  GCR originate from the wind material of massive {\it rotating}
stars, {\it always rich in CNO} (but
not in heavier nuclei); they are accelerated by the forward shock of the SN explosion.
}
\label{eps2}
\end{center}
\end{figure}

\section{Origin of cosmic rays}

For quite some time it was thought that GCR originate from the average ISM, where they
are accelerated by the {\it forward shocks} of SN explosions (Fig. 3.A). However, this can only produce secondary Be. 

A $\sim$constant abundance of C and O in GCR  can ``naturally" be understood if SN
accelerate their own ejecta, trough their {\it reverse schock} (Ramaty et al. 1997, see Fig. 3.B).  
However, the absence of unstable $^{59}$Ni 
(decaying through e$^-$ capture
within 10$^5$ yr) from observed GCR suggests that acceleration occurs 
$>$10$^5$ yr after the explosion (Wiedenbeck et al. 1999) 
when SN ejecta are presumably  already diluted in the ISM.  Furthermore, the reverse shock
has only a small fraction of the SN kinetic energy, while observed GCR require a large fraction
of it\footnote{The power of GCR is estimated to $\sim$10$^{41}$ erg s$^{-1}$ galaxywide, i.e.
about 10\% of the kinetic energy of SN, which is $\sim$10$^{42}$ erg s$^{-1}$ (assuming
3 SN/century for the Milky Way, each one endowed with an average kinetic energy of 1.5 10$^{51}$ ergs).}.

Higdon et al. (1998) suggested  that GCR are accelerated out of  {\it superbubbles} (SB) material  
(Fig. 3.C), enriched by the ejecta of many SN as to have a large 
and $\sim$constant metallicity. In this scenario, it is the forward shocks of SN 
that accelerate material ejected from other, previously exploded SN. Furthermore, it has been
argued that in such an environment GCR could be accelerated to higher energies 
 than in a single SN remnant (Parizot et al. 2004).  
That scenario has also been  invoked in order to explain the present day source isotopic 
composition of GCR (Binns et al. 2005, Rauch et al. 2009). 
Notice that the main feature of that composition, namely
a large $^{22}$Ne/$^{20}$Ne ratio, is explained as due to the contribution of winds from Wolf-Rayet (WR) stars
(e.g. Prantzos et al. 1987), and the SB scenario offers a plausible (but not unique) framework in bringing together
contributions from both  SN and WR stars.

\begin{figure}[t!]
\begin{center}
\includegraphics[angle=-90,width=0.95\textwidth]{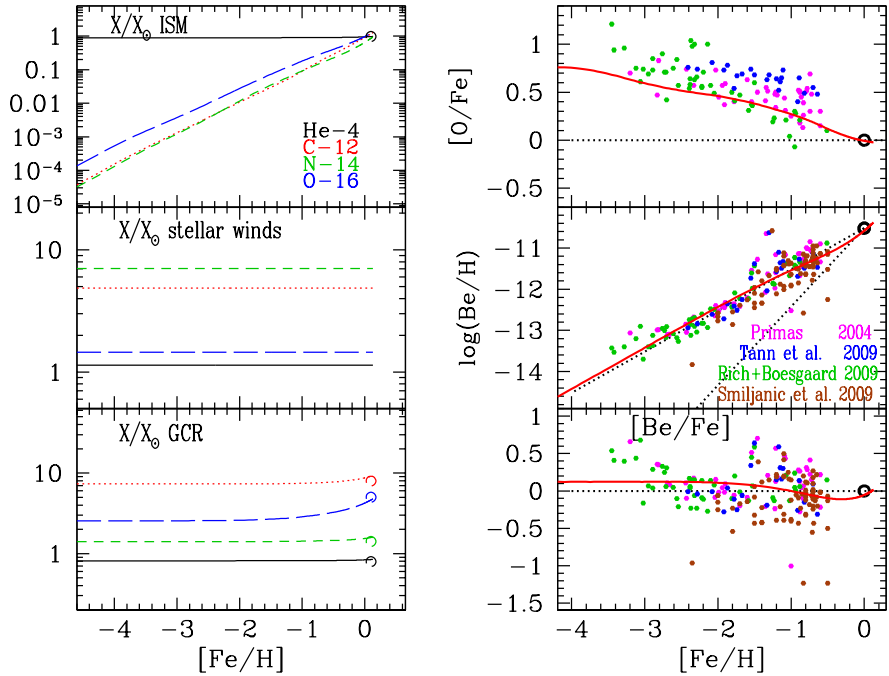}
\caption[]{{\it Left:} Evolution of the chemical composition (in corresponding 
solar abundances) of He-4 ({\it solid}), C-12 ({\it dotted}), N ({\it short dashed})
and O ({\it long dashed})in: ISM ({\it top}), massive star winds ({\it middle}) and
GCR ({\it bottom}). {\it Dots} in lower panel indicate estimated GCR source composition
(from Elison et al. 1997). {\it Right}: Evolution ({\it solid curves} of O/Fe ({\it top}),
Be/H ({\it middle}) and Be/Fe ({\it bottom}); {\it dotted lines} indicate solar values in
top and bottom panels, primary and secondary Be in middle panel.
}
\label{eps3}
\end{center}
\end{figure}

However, the SB scenario suffers from (at least) two problems.
First,  core collapse SN are observationally associated
to HII regions (van Dyk et al. 1996) and it is well known that the
metallicity of HII regions reflects the one of the {\it ambient ISM} (i.e.
it can be very low, as in IZw18) rather than the one of SN. 
Moreover, Higdon et al. (1998) evaluated the time interval $\Delta t$ between
SN explosions in a SB to a 
comfortable $\Delta t \sim$3 10$^5$ yr, leaving enough time to $^{59}$Ni 
to decay before the next SN explosion and subsequent acceleration. 
However,  Prantzos (2005) noticed that SB are constantly powered not only by SN
but also by the strong  winds of massive stars (with integrated energy and acceleration
efficiency similar to the SN one, e.g. Parizot et al. 2004), 
which should  continuously accelerate $^{59}$Ni, as soon as it is ejected from SN explosions.
Binns et al. (2008) argued that the problem may be alleviated from the fact that only the most massive
(and thus, short-lived) stars of an OB association emit strong winds; during the late (and longest)
 fraction of the lifetime of the SB 
 (a few 10$^7$ years) particles are accelerated episodically (by SN explosions only) and no more 
 continuously.
Still, it is hard to imagine that superbubbles have always the same average metallicity, especially during the
 early Galaxy evolution, where metals were easily expelled out of the shallow potential wells of the
 small sub-units forming the Galactic halo (e.g. Prantzos 2008).

\section{Cosmic rays from stellar winds and primary Be}

In this work we propose a different explanation for the origin of GCR, which can also provide
a satisfactory explanation for the primary nature of Be evolution. We first notice that there is now
substantial evidence that GCR are indeed accelerated in SN remnants (e.g. Berezhko et al. 2009 and references therein).
We then notice that, contrary to the case of non-rotating massive stars, 
which lose mass only at high metallicity, {\it rotating} massive stars display 
substantial mass loss down at very low (or even zero) metallicities (e.g.  Meynet, this volume).
The winds of those stars are enriched in CNO (products of H and He burning {\it within} the star itself)
at all metallicities and at about the same level; it is precisely this enrichment of the WR winds at
all metallicities that allows us to understand the observed primary behaviour of N down to the lowest
metallicity halo stars (Chiappini et al. 2006). This gives some confidence in using the same model results
to predict the composition of GCR over the history of the Milky Way.

 We assume then that GCR are accelerated when the forward shocks of SN propagate into the 
 previously ejected envelopes of rotating massive stars, which have been partially mixed with the
 surrounding ISM. The calculation of the resulting GCR composition $Y^{GCR}(M)$ is 
 far from trivial: it will be mostly $Y^{Wind}(M)$ in the case of SN with initial 
 mass $M>$20 M$_{\odot}$ (having lost a large fraction of their mass in the wind) and mostly 
 $Y^{ISM}$ in the case of $M$=10-20 M$_{\odot}$ stars, having suffered low mass losses. For ilustration
 purposes we adopt here, as a function of metallicity $Z$, 
 $Y^{GCR}_{paCNO}(Z)$=0.5 [$Y^{Wind}_{paCNO}(Z)$+$Y^{ISM}_{paCNO}(Z)$], where $Y^{Wind}(Z)$ is 
 provided by the Geneva models (G. Meynet, private communication) and is integrated over a stellar 
 IMF, whereas $Y^{ISM}(Z)$ is provided by the chemical evolution model (left panels in Fig. 4).

The calculation of the Be evolution is then
straightforward and nicely fits the data (right panels in Fig. 4); it is the first time that such a 
calculation is performed {\it not by assuming} a given $Y^{GCR}_{paCNO}(Z)$ but 
by {\it calculating} it in
a (hopefully) realistic way.

\begin{figure}[t!]
\begin{center}
\includegraphics[angle=-90,width=0.85\textwidth]{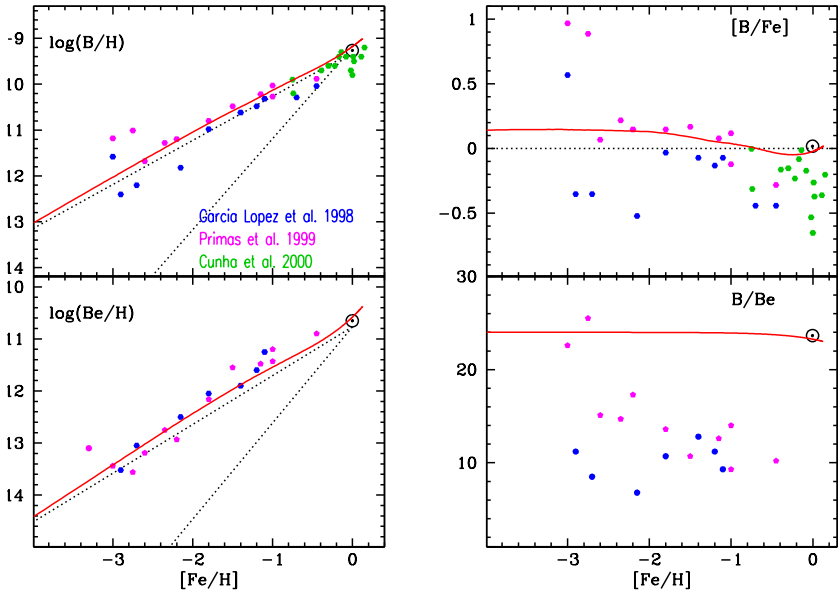}
\caption[]{{\it Left:} Evolution of B ({\it top}) and Be ({\it bottom}); in both panels,
{\it dotted lines} indicate primary and secondary evolution and {\it solid} curves indicate model evolution, including
apropriately normalised $\nu$-yields for $^{11}$B. {\it Right}:
Evolution of B/Fe ({\it top}) and B/Be ({\it bottom}). In the latter case, data indicate a subsolar mean
value of B/Be$\sim$14, compatible with exclusively GCR  production of both elements, but the uncertainties
(not shown here) are too large to allow conclusions.
}
\label{eps4}
\end{center}
\end{figure}

\section{Boron-11 from $\nu$-nucleosynthesis ?}

As mentioned in Sec. 2, a supplementary source of $^{11}$B is required in order to obtain the meteoritic
$^{11}$B/$^{10}$B=4 ratio. That source may be the $\nu$-process in SN,
 extensively studied in Woosley et al. (1990): a fraction of the most energetic among 
 the $\sim$10$^{59}$ neutrinos  of a SN explosion 
 spallate $^{12}$C nuclei in the C-shell of the 
 stellar envelope to provide $^{11}$B (but no other light nuclide). Soon after the HST observations
 of the primary behaviour of B (Duncan et al. 1992) it was realised that the $\nu$-process can 
 provide just such a primary B (Olive et al. 1994). But, if Be is produced as primary by GCR
  (Sec. 5), then more than $\sim$50\% of B is also produced as primary, leaving a rather small role 
  to the $\nu$-process. In fact,    the large uncertainties in the  $\nu$ 
 yields of $^{11}$B do not allow an accurate evaluation of the B evolution: 
 rather the B evolution (resulting
 from both GCR and $\nu$-process) has to be used in order to constrain the B yields of SN.
 
 The results of such an``exercise'' appear in Fig. 5. In order to fit the observations, the $\nu$ yields
 of Woosley and Weaver (1995) had to be divided by a factor of $\sim$6, otherwise B/H and B/Fe
  would be overproduced. Notice the model B/Be ratio is always $\sim$24 (i.e. solar), substantially higher
  than the observed, but {\it highly uncertain}, B/Be$\sim$14  ratio in halo stars (which is consistent with
  pure GCR production of both elements!). Clearly, future observations with HST are required to clarify
  that important issue.

\begin{figure}[t!]
\begin{center}
\includegraphics[angle=-90,width=0.85\textwidth]{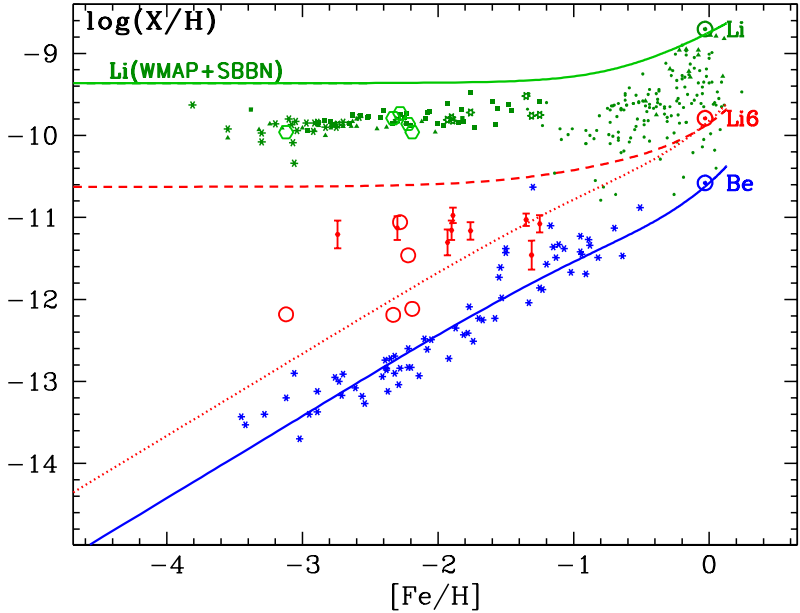}
\caption[]{Evolution of total Li ({\it upper} set of data points and {\it solid}
 curve for model assuming high primordial $^7$Li), 
Be ({\it lower} set of points and {\it solid} curve) 
and $^6$Li ({\it intermediate} set of points and curves). $^6$Li  data are from Asplund et al. (2006,
small filled circles with error bars) and Garcia-Perez et al. (2009, large open circles with - large - error bars not displayed), while
model curves are for a canonical (``low") pre-galactic $^6$Li ({\it dotted}) and a ``high" pre-galactic $^6$Li ({\it dashed}).   
In the latter case, a minimum amunt of depletion within stars (equal to that of $^7$Li) has been conservatively assumed.
}
\label{eps4}
\end{center}
\end{figure}

\section{Early $^7$Li  and $^6$Li: ``high'' or ``low'' ?} 

For a long time, the Li ``plateau" in low metallicity halo stars (discovered by Spite and Spite 1982) was considered to reflect the
primordial abundance of $^7$Li. However, the precise determination of baryonic density through observations of the
cosmic microwave background, combined to results of standard Big bang nucleosynthesis (SBBN), suggests that the
true value of primordial $^7$Li should be 2-3 times higher. It is not yet clear whether this discrepancy is due to some
problems with SBBN, whether non-standard particle physics might cure it, or whether primordial $^7$Li is depleted
in the surface convective zones of low metallicity stars with such an astonishing uniformity (see many contributions
in this volume). Other suggestions, like e.g. astration by a pre-galactic Pop. III population of massive stars (Piau et al. 2006)
face severe problems of metal overproduction (Prantzos 2006). This issue, one of the most important ones for
our understanding of mixing in stellar interiors, has also important implications for the chemical evolution of Li,
as we shall see below.

\begin{figure}[t!]
\begin{center}
\includegraphics[width=0.7\textwidth]{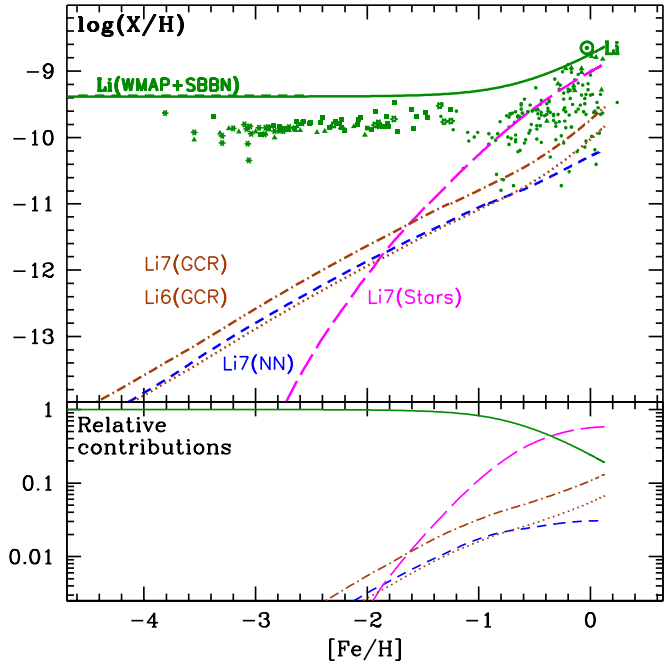}
\caption[]{Evolution of total Li ({\it top}) and percentages of its various components ({\it bottom}):
Li-7 from GCR ({\it dot-dashed}), Li-6 from GCR ({\it dotted}), Li-7 from $\nu$-nucleosynthesis 
(NN, {\it dashed}) and Li-7 from a delayed stellar source 
(novae and/or AGB stars, {\it long dashed}). {\it Solid} curves indicate total Li ({\it upper} panel)
and primordial$^7$Li ({\it lower} panel).
}
\label{eps4}
\end{center}
\end{figure}

The report of an ``upper envelope"  for $^6$Li/H in low metallicity halo stars
(Asplund et al. 2006) gave a new twist to the 
LiBeB saga. The reported $^6$Li/H value
at   [Fe/H]=-2.7 is much larger (by a factor of 20-30) than expected if GCR are the only source
of the observed $^6$Li/H in that star, assuming that GCR can account for the observed evolution of Be  (see Fig. 6).
But, if it turns out that the true primordial Li is the one corresponding to the WMAP+SBBN value, then the
initial $^6$Li  values in halo stars should be at least a factor of 3 higher than evaluated by Asplund et al. (2006, see Fig. 6).
It should be noticed, however, that such high $^6$Li values are not obtained in other investigations (Cayrel et al. 200,
Steffen et al. 2009).
 
 In the past few years, the possibility of important pre-galactic production of $^6$Li by non-standard GCR has drawn 
considerable attention from theoreticians, who proposed
several scenarios:

1) Primordial, non-standard, production during Big Bang Nucleosynthesis:
 the decay/annihilation of some massive particle (e.g.
neutralino) releases energetic nucleons/photons which produce $^3$He or  $^3$H
by spallation/photodisintegration of  $^4$He, while
subsequent fusion reactions between $^4$He and $^3$He or  $^3$H 
create  $^6$Li (e.g. Jedamzik 2004, and this meeting). Observations of  $^6$Li/H constrain
then the masses/cross-sections/densities of the massive particle. 

2) Pre-galactic, by fusion reactions of  $^4$He nuclei, accelerated
by  the energy released by massive stars (Reeves 2005) or by shocks induced 
during structure formation (Suzuki and Inoue 2002).

3) In situ production by stellar flares, through $^3$He+$^4$He reactions involving large amounts of accelerated $^3$He
(Tatischeff and Thibaud, 2007).

Prantzos (2006) showed that the energetics of $^6$Li production by accelerated particles constrain
severely any scenario proposed in category (2) above, including jets accelerated by massive black holes [this
holds  also for the ``stellar flare" scenario, the parameters of which have to be pushed to their extreme values
in order to obtain the ``upper envelope" of the Asplund et al. (2006) observations]. This difficulty is confirmed
by Evoli et al. (2008), who calculated pre-galactic $^6$Li production by $\alpha+\alpha$ reactions with a semi-analytical 
model for the evolution of the early Milky Way; they found maximum values shorter by factors $>$10 (and 
plausible values shorter by 3 orders of magnitude) than the values reported by Asplund et al. (2006).

\begin{figure}[t!]
\begin{center}
\includegraphics[angle=-90,width=0.7\textwidth]{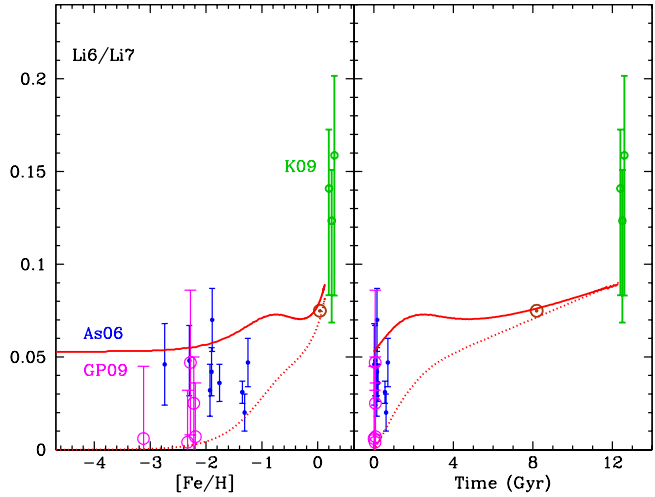}
\caption[]{{\it Left:} Evolution of Li6/Li7 ratio as a function of [Fe/H] ({\it left}) 
and of time ({\it right}). Data are from Asplund et al. (2006, As06), Garcia-Perez et al. (2009, GP09)
and Kawanomoto et al. (2009, K09). {\it Solid} curves corerspond to a ``high"
pre-galactic Li-6 and {\it dotted} curves to standard (low) pre-galactic Li-6. 
}
\label{eps5}
\end{center}
\end{figure}

\section{Evolution of Li and $^6$Li/$^7$Li}

Since GCR can only produce a $^7$Li/$^6$Li ratio of $\sim$2, instead of the
meteoritic (pre-solar) value of $\sim$12, another source of $^7$Li had to be found.
In the two decades following the original MAR paper, four such sources were identified:
three possible stellar sources and the hot early Universe of the Big Bang. The latter
has certainly operated, as testified by the observed Li ``plateau" in low metallicity halo 
stars; depending  on the true primordial value (see Fig. 7), it may contribute from 8 to 20\%
of the solar $^7$Li. Among the stellar sources, observational evidence exists
only for AGB stars, where high Li abundances have been detected in some cases.
But the corresponding model yields (from $^3$He+$^4$He in the bottom of the convective envelope)
are highly uncertain, and this is also the case for the other two candidate sources  of novae 
(from explosive H-burning) and core collapse SN (from $\nu$-induced nucleosynthesis); notice that
both novae and AGBs enter the Galactic scene with some time delay (``slow" $^7$Li component),
contrary to SN and GCR. 

$^7$Li is thus the only isotope having three distintinctively different types of sources: 
stellar, BBN and GCR. {\it Assuming} that the $\nu$-yields of $^7$Li are well established
(through the corresponding $^{11}$B yields, see Sec. 6), one may try to estimate
the evolution of the remaining ``slow" stellar contribution to $^7$Li, from the combined action
of novae and AGB stars, i.e. by removing from the observed evolutionary curve of Li/H vs Fe/H 
the BBN, GCR and $\nu$ contributions. The result
%of such an exercise 
is displayed in Fig. 7. The ``slow" stellar component contributes from
50-65\% of the solar $^7$Li (depending on whether high or low primordial $^7$Li is adopted);
similar numbers are found in the analysis of Matteucci (this volume).

Finally, Fig. 8 displays the evolution of $^6$Li/$^7$Li ratio, compared to data for the early halo
(highly uncertain, see previous section) and in the nearby Galactic disk (along three different lines 
of sight). Theoretical predictions depend on the adopted pre-galactic  $^6$Li/$^7$Li ratio, but
a generic feature is a late rise of $^6$Li/$^7$Li, due to the late secondary production of $^6$Li from GCR.

%\end{document}

\end{document}